\documentclass[prl,twocolumn,amsmath,amssymb,superscriptaddress]{revtex4}

\usepackage[pdftex]{graphics}
\usepackage{graphicx}
\usepackage{hyperref}

\renewcommand{\vec}[1]{{\mathbf #1}}
\newcommand{\rucl}{$\alpha$-RuCl$_3$}
\newcommand{\nair}{Na$_2$IrO$_3$}

\begin{document}

\title{Multiple Spin-Orbit Excitons and the Electronic Structure of $\alpha$-RuCl$_3$}

\author{P. Warzanowski}
\author{N. Borgwardt}
\author{K. Hopfer}
\author{T. C. Koethe}
\affiliation{Institute of Physics II, University of Cologne, 50937 Cologne, Germany}
\author{P.~Becker}
\affiliation{Sect. Crystallography, Institute of Geology and Mineralogy, University of Cologne, 50674 Cologne, Germany}
\author{V.~Tsurkan}
\affiliation{Experimental Physics V, Center for Electronic Correlations and Magnetism, University of Augsburg, 
	86135 Augsburg, Germany}
\affiliation{Institute of Applied Physics, MD 2028 Chisinau, Moldova}
\author{A.~Loidl}
\affiliation{Experimental Physics V, Center for Electronic Correlations and Magnetism, University of Augsburg, 
	86135 Augsburg, Germany}
\author{M. Hermanns}
\affiliation{Department of Physics, Stockholm University, AlbaNova University Center, SE-106 91 Stockholm, Sweden}
\affiliation{Nordita, KTH Royal Institute of Technology and Stockholm University, SE-106 91 Stockholm, Sweden}
\author{P. H. M. van Loosdrecht}
\author{M. Gr\"{u}ninger}
\affiliation{Institute of Physics II, University of Cologne, 50937 Cologne, Germany}

\date{November 21, 2019}

\begin{abstract}
The honeycomb compound \rucl\ is widely discussed as a proximate Kitaev spin-liquid material. 
This scenario builds on spin-orbit entangled $j$\,=\,1/2 moments arising for a 
$t_{2g}^5$ electron configuration with strong spin-orbit coupling $\lambda$ and a 
large cubic crystal field. 
The low-energy electronic structure of \rucl, however, is still puzzling. 
In particular infrared absorption features at 0.30\,eV, 0.53\,eV, and 0.75\,eV seem 
to be at odds with theory. Also the energy of the spin-orbit exciton, the excitation 
from $j$\,=\,1/2 to 3/2, and thus the value of $\lambda$ are controversial. 
Combining infrared and Raman data, we show that the infrared features can be attributed to 
single, double, and triple spin-orbit excitons. 
We find $\lambda$\,=\,0.16\,eV and $\Delta$\,=\,42(4)\,meV for the observed 
non-cubic crystal-field splitting, supporting the validity of the $j$\,=\,$1/2$ picture 
for \rucl. 
The unusual strength of the double excitation is related to the underlying hopping 
interactions which form the basis for dominant Kitaev exchange. 
\end{abstract}

\maketitle

The exactly solvable Kitaev model \cite{Kitaev06} describes bond-anisotropic exchange interactions 
on tricoordinated lattices. Exchange frustration embodied in the model yields a 
rich phase diagram with gapless and gapped quantum spin liquids in which spins fractionalize 
into emergent Majorana fermions and gauge fluxes. Gapless Majorana fermions form a metal 
which on a honeycomb lattice is equivalent to magnetic Dirac matter \cite{OBrien16}.  
Jackeli and Khaliullin \cite{Jackeli09} demonstrated that the Kitaev model may be realized in 
honeycomb compounds such as \nair\ or \rucl\ with edge-sharing IrO$_6$ or RuCl$_6$ octahedra, 
$t_{2g}^5$ configuration, and strong spin-orbit coupling. 
This triggered an avalanche of experimental and theoretical research 
\cite{Witczak14,Rau2016,Winter17rev,Trebst2017,Hermanns18,Takagi19}.

At low temperatures, \rucl\ orders magnetically due to exchange interactions 
beyond the Kitaev model \cite{Majumder15,Johnson15,Winter17}. 
The term {\em proximate Kitaev spin liquid} \cite{Banerjee16} was coined for \rucl\ 
based on fingerprints of dominant Kitaev interactions in spectroscopy, 
e.g., above the magnetic ordering temperature. 
In particular, inelastic neutron scattering, Raman scattering, and THz spectroscopy 
revealed an intriguing continuum of magnetic 
excitations \cite{Banerjee16,Banerjee17,Sandilands15,Nasu16,Glamazda17,Wang18,Little17,Wang17} 
for which a fermionic character \cite{Nasu16,Wang18} and the restriction of 
dynamical spin-spin correlations to nearest neighbors \cite{Banerjee17} were reported. 
However, a more conventional interpretation in terms of overdamped magnons has also been 
proposed \cite{Winter17}. 
The Kitaev picture is in line with the stunning claim of a half-quantized thermal Hall effect \cite{Kasahara18}.

In the light of these far-reaching results, it is surprising that the low-energy electronic structure 
of \rucl\ remains controversial. In the related iridates, the approximate validity of the 
$j$\,=\,1/2 scenario is well established by resonant inelastic x-ray scattering (RIXS) data of the 
spin-orbit exciton \cite{Kim12,KimNatComm14,Gretarsson13PRL,Rossi17,Revelli19a}, 
i.e., the excitation to $j$\,=\,$3/2$, but high-resolution RIXS at the Ru $L$ or $M$ edges is 
challenging \cite{Gretarsson19,Lebert19}. 
In \rucl, conflicting energies of 145\,meV, 195\,meV, and 231\,meV 
were reported for the spin-orbit exciton by Raman, 
inelastic neutron scattering, and Ru $M$ edge RIXS, respectively \cite{Sandilands16,Banerjee16,Lebert19}. 
Furthermore, several groups reported on infrared absorption bands at 
0.30\,eV, 0.53\,eV, and 0.75\,eV \cite{Binotto71,Guizzetti79,Plumb14,Sandilands16,Reschke17,Reschke18,Hasegawa17}. 
The 0.30\,eV peak was assigned to the Mott gap \cite{Plumb14,Reschke17,Hasegawa17} 
but a gap of 1.1\,eV is well established by different techniques
\cite{Binotto71,Guizzetti79,Pollini94,Kim15,Sandilands16,Sandilands16b,Koitzsch16,Koitzsch17,Biesner18}. 
Thus far, there is no convincing explanation for this multitude of excitations below the gap.

A scenario first proposed in 1971 attributes the infrared features to $t_{2g}^4\,e_g^1$ excited 
states \cite{Binotto71,Guizzetti79,Sandilands16,Hasegawa17,Biesner18}. 
This contradicts results of quantum chemistry and spectroscopy 
\cite{Yadav16,Koitzsch16,Agrestini17} which show the $t_{2g}^4\,e_g^1$ states above $1.3$\,eV.\@ 
Low-lying $e_g$ states would severely affect exchange interactions and the role of Kitaev 
coupling \cite{Jackeli09,WangHubbard17}. 
Alternatively, the infrared bands were discussed as possible evidence for a large non-cubic 
crystal-field splitting $\Delta$\,$\approx$\,$180$\,meV \cite{Reschke18}, 
which contrasts with the much smaller values found by x-ray absorption, 
$\Delta$\,$\approx$\,$-10$\,meV \cite{Agrestini17}, 
quantum chemistry, $\Delta$\,=\,$39$\,meV \cite{Yadav16}, 
density functional theory, $\Delta$\,=\,$37$\,meV \cite{Winter16,Wintervalue}, 
and $M$ edge RIXS, $\Delta < 40$\,meV \cite{Lebert19}.
Note that $\Delta$\,$\approx$\,$180$\,meV would strongly mix $j$\,=\,1/2 and 3/2 
states, again with dramatic consequences for the exchange interactions \cite{Chaloupka15,Winter17rev}. 
Finally, an exact diagonalization study interpreted the 0.3\,eV peak as the spin-orbit exciton, 
activated by direct $d$-$d$ intersite hopping \cite{KimShirakawa16}, 
leaving the features at 0.53 and 0.75\,eV unexplained.

\begin{figure}[t]
	\centering
	\includegraphics[width=0.9\columnwidth]{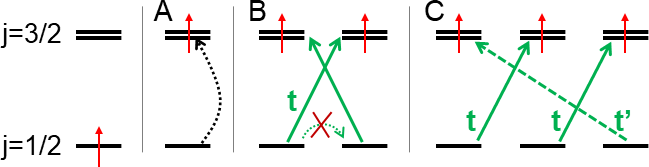}
\caption{{\bf Sketch of the different excitations} from a ground state with $j$\,=\,$1/2$ moments 
	on each site. The spin-orbit exciton, peak A in Fig.\ \ref{fig:IR_Raman}, is an 
	on-site excitation from $j$\,=\,$1/2$ to $3/2$. 
    A double spin-orbit exciton, peak B, results from the exchange of two holes between adjacent sites. 
	Nearest-neighbor hopping $t$ (solid green arrows) connects $j$\,=\,$1/2$ and $3/2$ on adjacent sites, 
	it is suppressed between $j$\,=\,$1/2$ states. 
	The triple process, peak C, involves hopping $t'$ between next-nearest neighbors. 
	The sketch neglects the non-cubic crystal field.
	}
	\label{fig:123sketch}
\end{figure}

In this letter, we report on infrared absorption and Raman scattering measurements 
which resolve the puzzle. Our Raman data feature the spin-orbit exciton, split by the 
non-cubic crystal field, at 248\,meV and 290\,meV.\@  
The three infrared bands correspond to phonon-assisted excitations of single and 
double spin-orbit excitons and the direct excitation of a triple spin-orbit exciton, 
see Fig.\ \ref{fig:123sketch}. 
We find that the phonon-assisted excitation of double spin-orbit excitons in Kitaev materials 
is closely related to the Lorenzana-Sawatzky-type two-magnon-plus-phonon absorption 
in the high-$T_c$ cuprate parent compounds \cite{Lorenzana95,Lorenzana95b}. 
Our data yield $\lambda$\,=\,0.16\,eV within the $4d$ shell. 
The ratio of non-cubic crystal-field splitting and $\lambda$ is very similar to the case of \nair, 
supporting the validity of the $j$\,=\,1/2 picture in \rucl.

Single crystals of \rucl\ were prepared by sublimation growth in evacuated SiO$_2$ glass ampoules 
after purification of the compound by recrystallization in vacuum.
Infrared transmittance was measured using a Bruker IFS\,66/v Fourier-transform 
spectrometer equipped with a $^4$He cryostat. We studied samples with a thickness of 
$(67 \pm 5)$\,$\mu$m and $(30 \pm 3)$\,$\mu$m. 
The polarization of the electric field was parallel to the honeycomb layers. 
We used the Fabry-Perot interference fringes to determine the refractive index $n$, which is 
approximately constant for frequencies below the Mott gap and far above the phonons. 
Knowing $n(\omega)$ and transmittance $T(\omega)$, one can calculate 
the optical conductivity $\sigma_1(\omega)$.  
Raman measurements were performed in back-scattering geometry using a micro-Raman setup with a 
TriVista spectrometer and an incident laser wavelength of 462\,nm.

\begin{figure}[t]
	\centering
	\includegraphics[width=0.94\columnwidth]{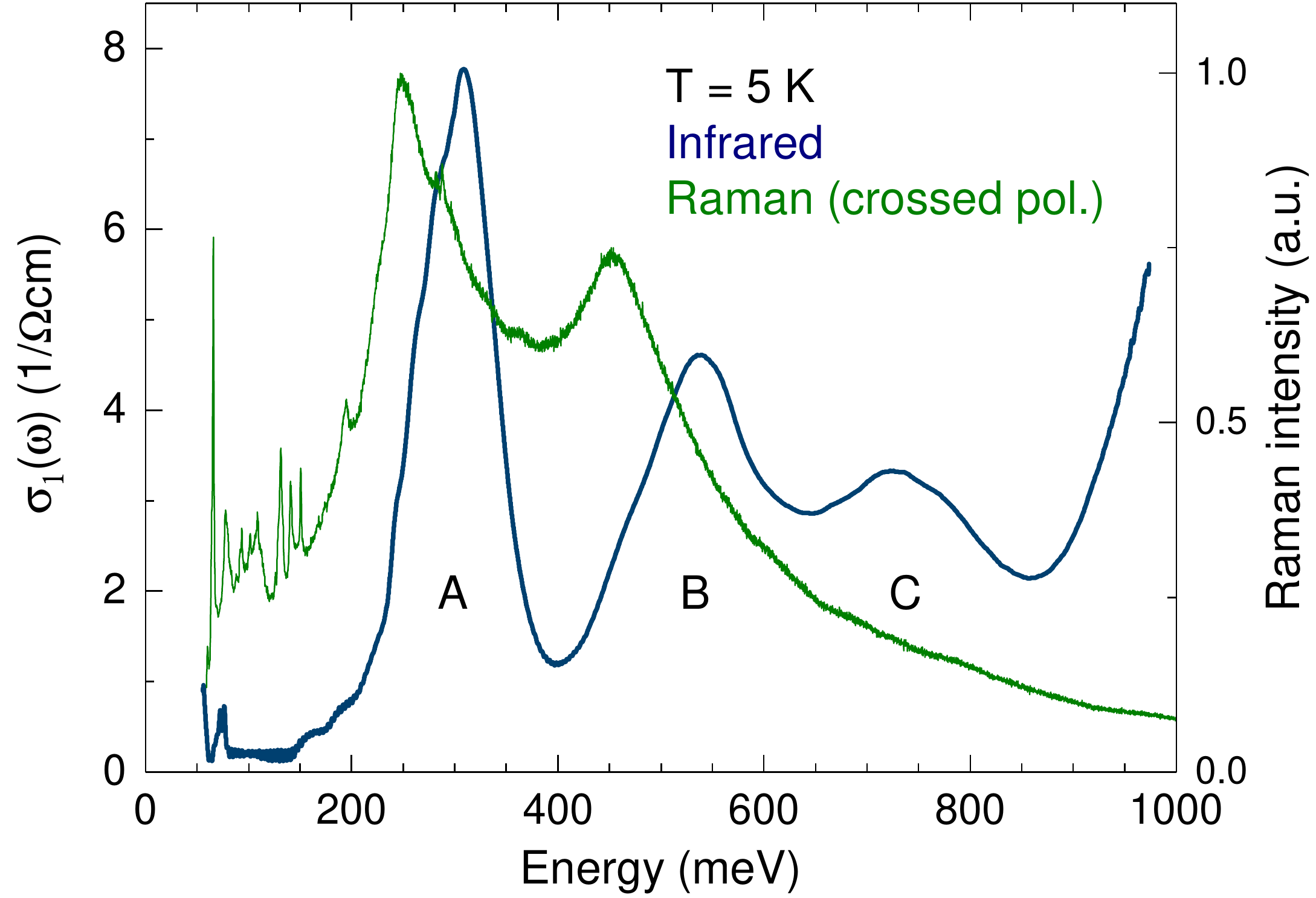}
	\caption{{\bf Optical conductivity $\sigma_1(\omega)$ (left axis) and Raman data (right) 
	of \rucl.} Features A, B, and C correspond to (phonon-assisted) 
	single, double, and triple spin-orbit excitons, respectively, cf.\ Fig.\ \ref{fig:123sketch}. 
	The increase of $\sigma_1(\omega)$ above 0.9\,eV indicates the Mott gap. 
	Above 1\,eV, $\sigma_1(\omega)$ reaches about $10^3$\,$(\Omega$cm$)^{-1}$ \cite{Sandilands16b,Biesner18}. 
	The much smaller values of $\sigma_1(\omega)$ below the gap are typical 
	for weak phonon-assisted or higher order absorption features \cite{Rueckamp}.  		
	}
	\label{fig:IR_Raman}
\end{figure}

We compare the optical conductivity $\sigma_1(\omega)$ and the Raman data in Fig.\ \ref{fig:IR_Raman}. 
In \rucl, phonons are restricted to below 40\,meV \cite{Sandilands15,Hasegawa17,Reschke18} 
and the energy of magnetic excitations is even smaller \cite{Banerjee16,Banerjee17,Sandilands15}. 
Above 0.9\,eV, $\sigma_1(\omega)$ shows the onset of excitations across the 
Mott gap, reaching about $10^3$\,$(\Omega$cm$)^{-1}$ above 1\,eV
\cite{Binotto71,Guizzetti79,Pollini94,Kim15,Sandilands16,Sandilands16b,Koitzsch16,Koitzsch17,Biesner18}. 
The prominent infrared peaks A, B, and C at 0.30\,eV, 0.53\,eV, and 0.75\,eV 
agree very well with previous 
reports \cite{Binotto71,Guizzetti79,Plumb14,Sandilands16,Reschke17,Reschke18,Hasegawa17}, 
demonstrating that these features reflect the local electronic structure. They are robust against 
the widely discussed sample issues in \rucl\ which are related to the stacking sequence of 
honeycomb layers \cite{Reschke18}. 
The Raman data show two strong peaks at 248\,meV and 450\,meV and a clear shoulder at 290\,meV.\@ 
The pronounced difference between infrared and Raman spectra originates from the selection rules,  
offering a key to the assignment.

A given mode is Raman active if it modulates the polarizability and infrared active if it carries an  
electric dipole moment. 
The spin-orbit exciton is a local excitation between $4d$ orbitals, see Fig.\ \ref{fig:123sketch}, 
closely related to an on-site $d$-$d$ excitation. 
Its Raman activity was demonstrated in Sr$_2$IrO$_4$ \cite{Yang15}. 
Such an electronic transition between even $d$ orbitals does not carry a dipole moment but becomes 
infrared-active by the simultaneous excitation of a symmetry-breaking phonon \cite{Henderson,Rueckamp}. 
Such phonon-assisted infrared excitations from $j$\,=\,1/2 to 3/2 were observed in host crystals 
doped with $5d^5$ Ir$^{4+}$ ions \cite{Yoo86}, equivalent to $4d^5$ Ru$^{3+}$.  
In the honeycomb lattice, the Ru sites do not show inversion symmetry. 
However, the corresponding admixture of $p$ or odd character to the $d$ orbitals usually is neglected. 
We thus expect the infrared peak of the spin-orbit exciton to be shifted with respect to the Raman peak 
by a phonon frequency, as observed for peak A.\@

\begin{figure}[t]
	\centering
	\includegraphics[width=0.94\columnwidth]{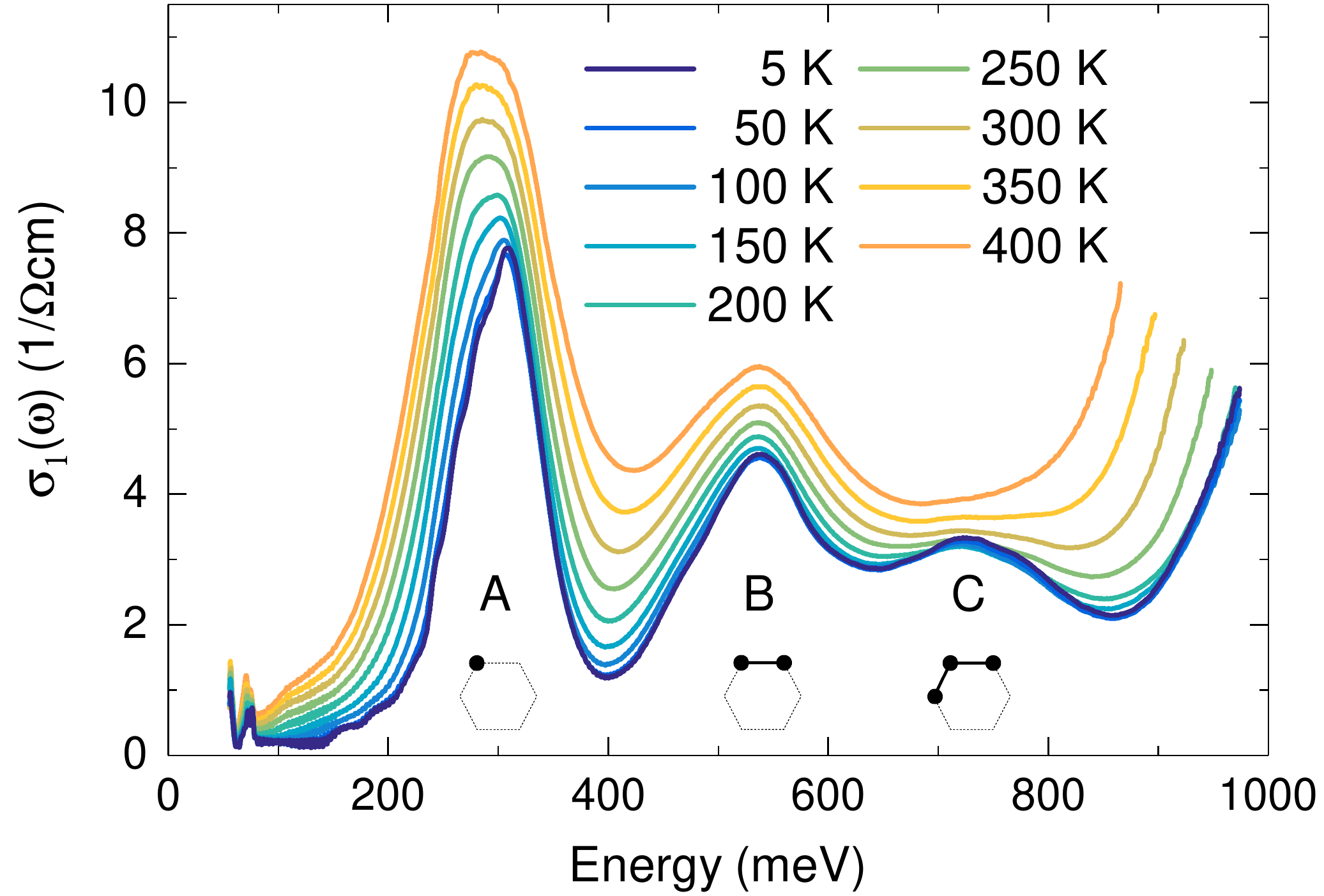}
	\caption{{\bf Temperature dependence of $\sigma_1(\omega)$.} 
		With increasing temperature, the Mott gap softens or smears out, partially covering peak C, 
		while peaks A and B show a pronounced increase of spectral weight, typical for phonon-assisted excitations, see Fig.\ \ref{fig:ab}a. 
		The high-energy cut-off reflects the suppression of the measured transmittance. 
		Sketches denote the different excitations, cf.\ Fig.\ \ref{fig:123sketch}.
	}
	\label{fig:sigma}
\end{figure}

\begin{figure}[t]
	\centering
	\includegraphics[width=\columnwidth]{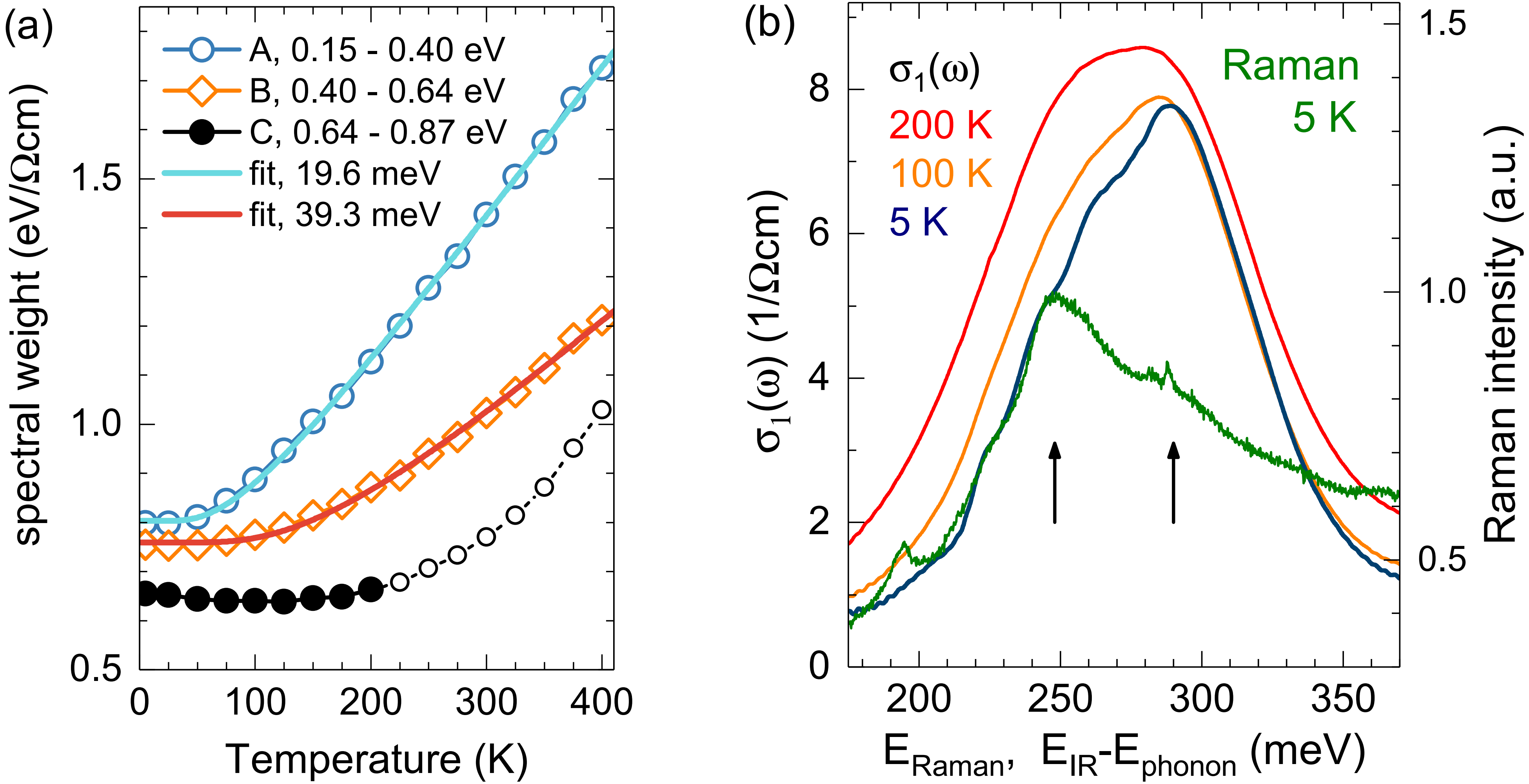}
	\caption{{\bf Spectral weight (left) and zoom in on peak A (right).} 
	Using Eq.\ \ref{eq:SW}, fits of the spectral weight, integrated over the indicated ranges,  
	yield $\hbar \omega_{\rm ph}$\,=\,19.6\,meV for peak A and 39.3\,meV for peak B.\@ 
	For peak C, the behavior above 200\,K (small symbols) is hidden by the softening 
	of the gap, cf.\ Fig.\ \ref{fig:sigma}, 
	while the spectral weight below 200\,K (full symbols) is independent of temperature. 
    Right panel: 
	Green: Raman data (right axis). 
	Other curves: $\sigma_1(\omega)$ (left axis), shifted by 20\,meV to compensate 
	for the phonon shift $\hbar \omega_{\rm ph}$ of the phonon-assisted process. 
	Arrows at 248\,meV and 290\,meV indicate peak positions of the spin-orbit exciton 
	split by the non-cubic crystal field. 
	Additional phonon sidebands reflect the vibronic character. 
}
\label{fig:ab}
\end{figure}

For phonon-assisted features located between $\omega_1$ and $\omega_2$, 
the phonon energy $\hbar \omega_{\rm ph}$ can be determined by the temperature dependence 
of the spectral weight \cite{Henderson}, 
\begin{equation}
SW = \int_{\omega_1}^{\omega_2} \sigma_1(\omega)\, d\omega = 
\alpha + \beta \coth\left(\frac{\hbar \omega_{\rm ph}}{2k_B T}\right) \, ,
\label{eq:SW}
\end{equation}
where $\alpha$ and $\beta$ are fit parameters. 
Thermal population of the phonon yields an increase of $SW$. 
Figure \ref{fig:sigma} depicts the temperature dependence of $\sigma_1(\omega)$, 
and Fig.\ \ref{fig:ab}a shows $SW$ of peaks A, B, and C.
The spectral weight of peak C is constant below 200\,K.\@  
The increase above 200\,K is due to the softening of the gap, 
see Fig.\ \ref{fig:sigma}. 
Accordingly, peak C is not phonon-assisted but directly infrared active, in agreement 
with the absence of a corresponding Raman feature. 
For peaks A and B, fits using Eq.\ (\ref{eq:SW}) yield $\hbar \omega_{\rm ph}$\,=\,19.6\,meV 
and 39.3\,meV, respectively, see Fig.\ \ref{fig:ab}a.

The phonon energies of 19.6 and 39.3\,meV agree with phonon studies using Raman 
and infrared spectroscopy \cite{Sandilands15,Hasegawa17,Reschke18}. 
The fact that two phonon energies are observed is remarkable. 
It is hard to reconcile with the scenario of excitations to various $t_{2g}^4\,e_g^1$ multiplets  
\cite{Binotto71,Guizzetti79,Sandilands16,Hasegawa17,Biesner18} because the phonon should be 
equally effective in breaking the symmetry for all on-site excitations. 
However, two different phonon modes have to be expected if we assign peaks A and B to 
single and double spin-orbit excitons. 
The former is an on-site process for which a phonon has to break the symmetry 
(on the honeycomb lattice: enhance the odd-symmetry character)
\textit{on a Ru site}. 
The double excitation, in contrast, involves the exchange of holes between two neighboring sites 
and becomes infrared-active via a different phonon mode breaking the symmetry 
\textit{between a Ru-Ru pair}.

\textit{Single spin-orbit exciton:} 
The closer look on peak A provided in Fig.\ \ref{fig:ab}b 
strongly supports its spin-orbit exciton nature. 
As discussed in the following, its energy yields a reasonable value of $\lambda$, 
and the data support the phonon shift between Raman and infrared data as well as the vibronic character 
expected for an excitation involving orbitals. 
Shifting the infrared data by 20\,meV to compensate for the phonon-assisted character, 
we find a stunning agreement with the Raman data which peak at 248\,meV and show a 
second feature at 290\,meV.\@ Both coincide with peaks in the shifted infrared data. 
The data show further sidebands shifted by 19\,meV.\@
These can be understood as vibronic sidebands according to the Franck-Condon principle \cite{Henderson,Rueckamp}.  
The $j$\,=\,1/2 ground state and the $j$\,=\,3/2 excited state differ in the spatial distributions of 
electronic charge and thus also in the relaxed configurations of the lattice \cite{Plotnikova16}. 
Therefore, these excitations mix with phonons which causes phonon sidebands. 
We emphasize that this vibronic mixture is a property of the eigenstates which is independent 
of the spectroscopic technique. In particular, the phonon sidebands should not be confused 
with the additional phonon that is necessary for breaking the symmetry in the infrared case.

The spin-orbit exciton can be described by the single-site Hamiltonian for the $t_{2g}$ 
states \cite{Jackeli09,Sizyuk14}, 
\begin{equation}
H_{\rm single} = \lambda_{\rm eff}\, \vec{S}\cdot \vec{L} + \Delta_{\rm CF}\, L_{[111]}^2 \, ,  
\label{eq:single}
\end{equation}
where $L_{[111]}$ denotes the component of $\mathbf{L}$ along the trigonal $[111]$ axis. 
The trigonal crystal field parameterized by $\Delta_{\rm CF}$ yields a splitting of the excited 
$j$\,=\,$3/2$ quartet.
For $\Delta_{\rm CF}/\lambda_{\rm eff}$\,$\ll$\,$1$, the observed splitting is 
$\Delta$\,=\,$\frac{2}{3}\Delta_{\rm CF}$. 
With peaks at 248(1)\,meV and 290(4)\,meV, there are two different solutions of Eq.\ (\ref{eq:single}), 
$\lambda_{\rm eff}$\,=\,175(1)\,meV and $\Delta_{\rm CF}$\,=\,$70(9)$\,meV 
for trigonal elongation or
$\lambda_{\rm eff}$\,=\,177(1)\,meV and $\Delta_{\rm CF}$\,=\,$-59(7)$\,meV
for compression. 
Accordingly, the $j$\,=$1/2$ wavefunction carries more than 98\,\% of the weight of the local 
ground state, corroborating the applicability of the $j$\,=\,1/2 picture for \rucl. 
The ratio $\Delta/\lambda_{\rm eff}\! \approx \! 1/4.5$ is very similar in \nair, 
where RIXS finds $\lambda_{\rm eff}$\,=\,0.4--0.5\,eV and $\Delta$\,=\,0.11\,eV \cite{Gretarsson13PRL}.

Quantum chemistry calculations predicted the spin-orbit exciton at 195\,meV and 234\,meV 
with a splitting $\Delta$\,=\,39\,meV \cite{Yadav16}. 
These values, in particular the splitting, agree very well with our result. 
Density functional theory finds $\Delta$\,=\,$37$\,meV \cite{Winter16,Wintervalue}
while a somewhat smaller splitting of $\Delta_{\rm CF}$\,=\,$(-12\pm 10)$\,meV 
was derived from linear dichroism in x-ray absorption spectroscopy \cite{Agrestini17}. 
Note that our experimental splitting $\Delta$\,=\,$42(4)$\,meV is still smaller 
than the vibronic peak width, and one may speculate whether the considerable overlap 
of the two excitations explains the reduced dichroism. With the currently available resolution, 
RIXS data measured at the Ru $M$ edge at 300\,K support our assignment of peak A 
but cannot resolve the splitting \cite{Lebert19}.

The effective parameter $\lambda_{\rm eff}$\,$\approx$\,176\,meV applies to the single-site Hamiltonian 
restricted to $t_{2g}$ orbitals, Eq.\ (\ref{eq:single}). Considering the full $4d$ shell including 
the $e_g$ orbitals with 10\,Dq\,=\,2.2\,eV \cite{Koitzsch16,Sandilands16}, 
we find $\lambda$\,=\,0.16\,eV.\@ 
In the literature on \rucl, typical values for $\lambda$ fall in the range 
0.13--0.15\,eV \cite{Kim15,Agrestini17,Sinn16,WangHubbard17,Lebert19}, 
but a precise experimental value was missing thus far.

Our interpretation of peak A as spin-orbit exciton poses the question of the origin 
of the features at lower energy that previously were attributed to this excitation. 
The claim of a neutron mode at 195\,meV \cite{Banerjee16} needs to be substantiated by 
high-energy neutron data with improved signal-to-noise ratio.
In Raman scattering with a 532\,nm laser, a weak mode was reported at 145\,meV \cite{Sandilands16}.
Our Raman data show a multitude of weak modes below 200\,meV, see Fig.\ \ref{fig:IR_Raman}.
In particular we find three narrow peaks at 131\,meV, 141\,meV, and 151\,meV.\@ 
We tentatively attribute all of these weak features to multi-phonon excitations. 
Their strength depends on the laser wavelength and the sample quality and may be boosted 
by a flat phonon dispersion. 
In the infrared data, the feature at 73\,meV, i.e., about half the energy, has been 
explained as a strong two-phonon excitation \cite{Hasegawa17}.

\textit{Double spin-orbit exciton:} 
To explain this process, we consider the phonon-assisted two-magnon absorption described by 
Lorenzana and Sawatzky \cite{Lorenzana95,Lorenzana95b}, a powerful tool for the study of 
magnetic excitations in, e.g., the $S$\,=\,1/2 cuprate parent compounds \cite{Grueninger03}. 
There, exchanging two antiparallel spins on adjacent sites is equivalent to a double spin flip. 
This process is Raman active but does not carry a dipole moment. It becomes infrared active 
by the simultaneous excitation of a high-energy breathing phonon which yields different 
on-site energies on the two sites, breaking the symmetry on the bond. 
In the edge-sharing geometry of \rucl, this role is played by the highest $E_u$ phonon 
mode \cite{Hasegawa17}, which agrees with our result $\hbar\omega_{\rm ph}$\,=\,39\,meV 
for peak B.\@
The spectral weight in $\sigma_1(\omega)$ is proportional to the effective charge 
$q_{\rm eff}$\,=\,$\frac{\partial^2 J}{\partial u \, \partial |\mathbf{E}|}$  
where $J$ denotes the exchange coupling constant, $u$ the ionic displacement, 
and $\mathbf{E}$ the applied electric field. 
For \rucl, we estimate the order of magnitude of the leading-order contribution \cite{estimateqeff}, 
which yields similar results as in the cuprates. 
Experimentally, peak B reaches about 4\,$(\Omega$cm$)^{-1}$ and has a width of about 0.1\,eV.\@ 
Indeed, similar spectral weights are found in the cuprates \cite{Grueninger03}.

Compared to magnetic excitations, the observation of a double excitation involving orbitals 
is unusual but not unprecedented. The strength of this process depends on the particular form 
of intersite hopping. For the two-magnon process in the cuprates, it is sufficient to consider 
a half-filled $x^2$-$y^2$ orbital per site and hopping between them. 
In contrast, the double orbital excitation is boosted if hopping predominantly connects 
the ground state orbital on site $i$ with an excited orbital on site $j$. 
For instance, two-orbiton infrared absorption was observed in the Mott insulator 
YVO$_3$ \cite{Benckiser08} which shows antiferro orbital order of $xz$ and $yz$ along $c$. 
Hopping between adjacent sites $i$ and $j$ is diagonal in $xz$ and $yz$. 
Starting from $|0\rangle$\,=\,$|xz\rangle_i\,|yz\rangle_j$, the exchange of electrons leads 
to the final state  $|yz\rangle_i\,|xz\rangle_j$ with orbital excitations on both sites. 
In \rucl, the double excitation proceeds via a similar path. 
Nearest-neighbor hopping from $|j$\,=\,$1/2\rangle_i$ is only finite to $|3/2\rangle_j$ 
but vanishes to $|1/2\rangle_j$, see Fig.\ \ref{fig:123sketch}. 
From $|0\rangle$\,=\,$|1/2\rangle_i\,|1/2\rangle_j$, 
the leading contribution of particle exchange yields $|3/2\rangle_i|3/2\rangle_j$, 
the double spin-orbit exciton. Interaction effects change this picture only slightly, 
the double spin-orbit exciton remains the dominant contribution.
This property of the hopping interactions is the source for vanishing Heisenberg 
and dominant Kitaev exchange \cite{Jackeli09}. In this sense, the strength of the double 
excitation is a direct consequence of dominant Kitaev exchange.

The precise energy of the double excitation is more subtle. Naively, one may expect combination tones 
of 248\,meV and 290\,meV.\@ However, for the double excitation one has to consider dispersion, 
even if the total wavevector has to vanish, and in particular interaction effects. 
For two-magnon absorption, this dramatically affects the peak energies, in particular 
in the case of bound states \cite{Windt01}. 
In \rucl, it is, e.g., feasible to consider a lattice distortion which lowers the spin-orbit excitation 
energy on \textit{both} sites. The observed energies of 450\,meV in Raman scattering 
and (490+39)\,meV in infrared absorption are thus plausible. Following Fig.\ \ref{fig:123sketch}, 
the infrared peak invokes excitations to $|3/2\rangle_i|3/2\rangle_j$ due to nearest-neighbor hopping $t$. 
Raman scattering proceeds via a high-energy intermediate state which may favor a different flavor 
of the double spin-orbit exciton, explaining the lower energy. 
A thorough investigation of interaction effects is beyond the scope of our study, which is the first 
report of a double spin-orbit exciton. In the related iridates, 
RIXS at the Ir $L$ edge strongly favors on-site excitations such as the single spin-orbit exciton. 
However, a close look at the RIXS data of Na$_2$IrO$_3$ reveals a small feature at about twice 
the energy of the spin-orbit exciton \cite{Gretarsson13PRL}.

\textit{Triple spin-orbit exciton}: 
A triple spin-orbit exciton excitation is possible on three neighboring sites $i$, $j$, $k$ 
of the honeycomb lattice. It results from nearest-neighbor hopping $t$ from $i$ to $j$ and 
from $j$ to $k$ and next-nearest neighbor hopping $t'$ from $k$ to $i$, see Fig.\ \ref{fig:123sketch}. 
Following the logic of Lorenzana-Sawatzky-type phonon-assisted absorption \cite{Lorenzana95,Lorenzana95b}, 
the different hopping amplitudes in combination with the absence of inversion symmetry for this 
group of three sites result in a finite dipole moment, i.e., 
the excitation is directly infrared active.

A triple excitation is unusual as the spectral weight strongly decreases with 
increasing particle number. In the present case, this is partially compensated by the fact that 
a directly infrared-active process such as the triple excitation is much stronger than 
phonon-assisted processes like the single and double excitations. 
Moreover, next-nearest neighbor hopping is of considerable size in \rucl, 
which partially is caused by the strong hopping between the large Cl ions. 
Band-structure calculations find $t'/t \approx 0.38$ \cite{Winter16}.

In conclusion, we solved a long-standing puzzle concerning the low-energy electronic structure of \rucl. 
The prominent features below the Mott gap can be identified as single, double, and 
triple spin-orbit excitons. 
In \rucl, the spin-orbit exciton is far below the gap, which allows Raman scattering and 
infrared absorption to take over the important role that RIXS plays in iridates.  
We determine $\lambda$\,=\,0.16\,eV, a central parameter for the theoretical modeling of \rucl, 
and $\Delta$\,=\,42(4)\,meV, corroborating the $j$\,=\,1/2 scenario. 
The observation of a double spin-orbit exciton highlights the prominent role of Kitaev coupling 
and calls for studies of the interactions between excited $j$\,=\,3/2 states in Kitaev materials.

\begin{acknowledgments}
We acknowledge funding from the Deutsche Forschungsgemeinschaft (DFG, German Research Foundation) 
-- Project number 277146847 -- CRC 1238 (projects A02, B02, and B03). 
\end{acknowledgments}


\begin{thebibliography}{99}
\bibitem{estimateqeff}
In leading order one finds 
$q_{\rm eff}$\,$\propto$\,$e \, a \, \frac{J}{\Delta_{is}^2}\frac{\partial \Delta E_{M}}{\partial u}$, 
where $e$ denotes the electric charge, $a$ the metal-metal distance, 
$\Delta_{is}$ the effective inter-site excitation energy, and $\Delta E_{M}$ the phonon-induced difference 
in metal on-site energies. In the cuprates, exchange is an order of magnitude larger than in \rucl\ 
and $\Delta_{is}$ is about a factor of 2 larger. We thus find that $q_{\rm eff}$ is of the same order 
of magnitude.
\end{thebibliography}
\end{document}